\documentclass[12pt]{article}

\begin{document}

\sloppy
\begin{flushright}{SIT-HEP/TM-15}
\end{flushright}
\vskip 1.5 truecm
\centerline{\large{\bf Hybridized Affleck-Dine baryogenesis}} 
\vskip .75 truecm
\centerline{\bf Tomohiro Matsuda
\footnote{matsuda@sit.ac.jp}}
\vskip .4 truecm
\centerline {\it Laboratory of Physics, Saitama Institute of
 Technology,}
\centerline {\it Fusaiji, Okabe-machi, Saitama 369-0293, 
Japan}
\vskip 1. truecm
\makeatletter
\@addtoreset{equation}{section}
\def\theequation{\thesection.\arabic{equation}}
\makeatother
\vskip 1. truecm

\begin{abstract}
\hspace*{\parindent}
We propose a novel scenario for Affleck-Dine baryogenesis in
the braneworld, considering the hybrid potential for the Affleck-Dine field.
Destabilization of the flat direction is not due to the
Hubble parameter, but is induced by a trigger field.
The moduli for the brane distance plays the role of the trigger field.
Q-balls are unstable in models with large extra dimensions. 
\end{abstract}

\newpage
\section{Introduction}
\hspace*{\parindent}
In spite of the great success in the quantum field theory, there is still no
consistent scenario in which quantum gravity is successfully included.
Perhaps the most promising scenario in this direction would be the string
theory, in which the consistency is ensured by the
requirement of the additional dimensions.
In the old scenarios where no branes are included, sizes of the extra
dimensions had been assumed to be as small as $M_p^{-1}$.
However, later observations showed that there is no reason
to believe such tiny compactification radius\cite{Extra_1}.
In models with large extra dimensions, 
the observed Planck mass is obtained by the relation $M_p^2=M^{n+2}_{*}V_n$,
where $M_{*}$ and $V_n$ denote the fundamental scale of gravity
and the volume of the $n$-dimensional compact space.
In this respect, what we had seen in the old string theory was a tiny 
part of the whole story.
In the new scenario, the compactification radius (or the fundamental
scale) is the unknown parameter that should be determined by
observations.  
Until recently, cosmological models for such large compactification
radius had not been discussed.\footnote{Constructing successful models for
inflation with the low fundamental scale is an interesting
problem\cite{low_inflation, matsuda_nontach}.  
Baryogenesis in models with the low fundamental scale is discussed in
ref.\cite{low_baryo, low_AD, ADafterThermal}.}  
Constructing models for the particle cosmology with large
extra dimensions is very important since future cosmological
observations will play important roles in determining the underlying theory.

In this paper we consider novel ways to realize Affleck-Dine
baryogenesis\cite{AD} in models 
with the fundamental scale that is much lower than the conventional GUT scale.
Of course, there are drastic changes from the conventional scenarios.
In the conventional scenarios of Affleck-Dine baryogenesis, it had been found
that the dynamics of the Affleck-Dine field is rather complicated
because of the non-trivial formation and decay of
Q-balls\cite{Q-ball_kusenko}.  
In generic cases, almost all the produced baryon number is absorbed into
Q-balls.
Their properties depend on how supersymmetry breaking is transmitted.
If supersymmetry breaking is mediated by gravity, Q-balls are
semi-stable but long-lived and may be the source of all the baryons and
 LSP dark matter\cite{Problem_of_Q-ball}.
For the gauge-mediated scenarios, Q-balls can be stable and
form dark matter that can be searched for directly.   
Unstable Q-balls can decay to provide all the baryonic charges trapped
inside Q-balls, if some requirements are fulfilled.
If Q-balls are absolutely stable, it is difficult to produce the sufficient
baryon in the plasma although huge baryon number will be
kept inside Q-balls.
It is also a problem if Q-balls decay late to produce unwanted relics in
the Universe.
On the other hand,
when one considers models with the low fundamental scale, the gravitino mass
($m_{3/2}$) is much smaller than the soft breaking mass ($m_{soft}$)  
in the sector of the minimal supersymmetric standard model (MSSM).
In this respect, the situation looks similar to the gauge-mediated
supersymmetry breaking (GMSB) scenarios.
In GMSB models, however, the flat potential for the Affleck-Dine mechanism
becomes very flat at large amplitudes.
It is schematically given by the formula
\begin{eqnarray}
V_{GMSB} & \sim & m_{soft}^2 |\phi_{AD}|^2 \, \, \, \,
 (\phi_{AD}<<\Lambda_m), \nonumber\\
&\sim & V_0 \log \frac{|\phi_{AD}|^2}{\Lambda_m^2}\, \, \,
 (\phi_{AD}>>\Lambda_m)
\end{eqnarray}
where $\Lambda_m$ is the messenger scale.
Of course the gravitational effect always exists, which lifts the flat
direction by
\begin{eqnarray}
V_{GRA} &\sim& m_{3/2}^2|\phi_{AD}|^2.
\end{eqnarray}
Here the soft breaking mass $m_{soft}$ is much larger than the gravitino
mass $m_{3/2}$.
As the potential for the Affleck-Dine field is very flat at the large
amplitude, the Q-ball formation is inevitable.
In the gauge-mediated SUSY breaking models, the Q-ball formation makes
the scenario difficult to produce sufficient baryon number of the
universe while evading the cosmological problems.\footnote{There are many
discussions on this topic\cite{Problem_of_Q-ball,dineT}.}

As we will discuss in the followings, the above-mentioned problems
for the single-field models are naturally solved or modified 
in hybrid models.
Here we briefly show the basics of the idea.
In the conventional scenarios of Affleck-Dine baryogenesis, one should
assume $H>m_{\phi_{AD}}$ before the time of baryogenesis 
so that the flat directions are destabilized by the gravitational
corrections of $O(H)$, to obtain the large initial
amplitude of the baryon-charged directions.
This simple idea seems to solve the problem of the initial condition
in the original model\cite{AD}.
On the other hand, however, this simple condition sometimes puts a severe
constraint on the models, as we will discuss in the next section. 
To find a solution to the problem, we think it is interesting to invoke
ideas that had been used to solve the problems in other cosmological scenarios.
For example, when one considers chaotic inflation, problems arise for the
original single-field model.
The problems are solved by hybrid inflation, in which an additional field
is included to lift the energy density and trigger the termination of
inflation. 
Then it seems natural to ask,
``Is it possible to find the trigger field in Affleck-Dine baryogenesis,
which removes the serious constraints in the original single-field
models?'' 
In this paper we find the solution to the above (rather naive) question.
In generic situations, it seems quite hard to find hybrid models in the
conventional  
settings of supergravity.
Thus we consider the scenario for the braneworld where the moduli for
the brane distance can play the required role of the trigger
field\cite{matsuda_nontach}.
In this paper we focus our attention to the models with the low
fundamental scale.
The models with the low fundamental scale are attractive since the
cosmological scenarios are expected to become quite
different from the 
conventional one, which may leave distinguishable 
signatures in the present Universe.
The Q-balls are naturally unstable in the models with large
extra dimensions.

\section{Hybridized Affleck-Dine baryogenesis}
\hspace*{\parindent}
In this section we consider models for the braneworld in which the
fundamental scale ($M_*$) is much smaller than the conventional GUT
scale.
Naively, the model should be similar to the GMSB models, in which the
gravitino mass is much smaller than the conventional soft mass, and the
messenger scale is a cutoff scale for the effective theory.
The crucial difference is the initial condensate of the Affleck-Dine field,
which cannot become larger than the cutoff scale since the flat direction
is localized on the brane.
Of course in the conventional GMSB models,
it becomes much larger than the messenger scale.
As a result, flat potentials in GMSB models are destabilized until
$H\sim m_{3/2}<<m_{soft}$.
On the other hand, in models with the low fundamental scale,
the Affleck-Dine field starts to oscillate at $H\sim m_{soft}$
with the amplitude $<\phi_{AD}> \le M_*$ unless there is the non-trivial
mechanism to destabilize the potential. 
As is discussed in ref.\cite{low_AD}, the situation is hopeless because
the resultant baryon to entropy ratio is at most
\begin{eqnarray}
\frac{n_b}{s}&\simeq& \frac{n_b}{n_{\phi_{AD}}}
\frac{T_R}{m_{\phi_{AD}}}\frac{\rho_{\phi_{AD}}}{\rho_{I}}\nonumber\\
&<&\frac{T_R}{m_{\phi_{AD}}}\left(
\frac{m_{\phi_{AD}}^2 M_*^2}{m_{\phi_{AD}}^2 M_p^2}\right)\nonumber\\
&\sim& 10^{-29}\left(\frac{M_*}{10^6 GeV}\right)^2
\end{eqnarray}
for the reheating temperature $T_{R} \sim 1-10$MeV.
Here $\rho_{I}$ denotes the energy density of the inflaton field.
To avoid this difficulty, one should consider non-trivial realization of
the Affleck-Dine mechanism.\footnote{In ref.\cite{low_AD}, the
Affleck-Dine field is put into the bulk where
the large volume factor enhances the energy density of the Affleck-Dine
field.
In ref.\cite{ADafterThermal}, Affleck-Dine
baryogenesis after thermal brane inflation\cite{thermal_brane} was
considered.
We have also discussed the effect of the cosmological defects in
ref.\cite{low_baryo}.}  

To avoid the above difficulty,
we show the non-trivial mechanism to destabilize the flat direction on
the brane. 
Here we consider the case in which the F-term on a brane destabilizes
the Affleck-Dine flat direction.
As in the models for brane inflation, we consider two branes separated
at a distance.
At the beginning of Affleck-Dine baryogenesis, these two branes are
required to be located at a distance in the extra dimensions.
We assume that in the true vacuum, the moduli for the brane distance is
stabilized by the mass of $O(m_{3/2})$, while it is destabilized by the
O(H) gravitational corrections during inflation.
Then the moduli is destabilized until $H \sim m_{3/2}$, which means that
two branes are separated during this period.
On one brane, for the simplest example, we assume the localized field S
and the superpotential of 
the form\cite{matsuda_nontach} 
\begin{equation}
W_1=S\Lambda_1^2
\end{equation}
which spontaneously breaks supersymmetry by the F-term.
On the other brane, a superfield $\Phi$ is localized with the
superpotential $W_2=0$. 
When two branes are on top of each other, the localized fields $S$ and
$\Phi$ may interact. 
Then finally in the true vacuum, where two branes are on top of each
other, a superpotential appears on the brane, 
\begin{equation}
W_{1+2}=S(\Lambda_1^2-\Phi^2),
\end{equation}
which restores supersymmetry.
One may expect many other forms of the
superpotential\cite{matsuda_nontach}.
The requirement for the mechanism is very simple.
Supersymmetry is needed to be spontaneously broken when the interactions
are absent, while it is restored in the vacuum. 
If the supersymmetry breaking terms induced by the above simple
mechanism dominate the potential for the Affleck-Dine field, the
Affleck-Dine field on the brane is destabilized.
In this case, the required trigger field is the moduli field that
parametrizes the distance between branes.

In the above example, the oscillation of the Affleck-Dine field 
starts at $H\sim m_{3/2}<<m_{soft}$.
Since the oscillation starts much later than the conventional
single-field models, the baryon to entropy ratio can be
enhanced in our model.
In the most optimistic case, when $\rho_{AD} \sim
m_{\phi_{AD}}^2(\phi_{AD}^i)^2$, 
\begin{equation}
\frac{n_{B}}{s}\sim \frac{T_{R}m_{\phi_{AD}}^2
(\phi_{AD}^i)^2}{m_{\phi_{AD}} \rho_I}
\end{equation}
where $T_{R}$ is the reheating temperature after Affleck-Dine
 baryogenesis, and $\phi_{AD}^i$ is the initial amplitude of
$\phi_{AD}$.
Then we obtain:
\begin{equation}
\frac{n_{B}}{s}\sim 10^{-9} \left(\frac{T_{R}}{10MeV}\right)
\left(\frac{(10^{5} GeV)^4}{\rho_I}\right)
\left(\frac{\phi_{AD}^i}{10^6 GeV}\right)^2
\end{equation}
which is the most naive result, but is enough to explain the origin of
the baryon asymmetry of the present Universe.
The Hubble parameter when the AD oscillation starts is assumed to be
 $H_o \simeq m_{3/2}$. 
It is naturally assumed that the initial amplitude is as large as
$\phi_{AD}^{ini}\sim M_*$.
To be more precise, the baryon to entropy ratio is determined by the
 forms of the A-terms, which explicitly break baryon number\footnote{
See ref.\cite{AD_dine} for more detail.
Here we had not specified the phenomenological mechanism for
supersymmetry breaking and its mediation, because we are not making a
catalog in this paper.
Although there are so many models in this direction, they are
not directly related to the characteristic profiles of our model. 
Of course the produced baryon to entropy ratio depends on the
A-terms.
When the fundamental scale is much lower than the GUT scale, one should
consider some non-trivial mechanisms to suppress the dangerous
higher-dimensional terms that explicitly break the baryon number
conservation. 
There are many discussions about the mechanism for suppressing the
dangerous operators, which in turn determines
the forms of the required A-terms.
However, here we do not consult into the details
 of the arguments but simply assume that there is the required
A-term when Affleck-Dine baryogenesis starts.
Of course, if the required A-terms are completely forbidden by some
 gauged symmetries, it is impossible to produce any baryon number by
 using the Affleck-Dine mechanism. 
In ref.\cite{low_baryo}, we have already considered
models in which the baryon number violating interactions are enhanced by
the cosmological defects.}.

It is easy to see that our mechanism works for the
models with the large fundamental scale (or small extra dimensions).
In these models, however, there is no compelling motivation to consider hybrid
models since (at this time) there seems no obvious advantage of the
hybrid model. 
We think our model is stringent for the models with the low fundamental scale,
where the conventional scenario utterly fails\cite{low_baryo, low_AD,
ADafterThermal}. 

\section{Inverted scenarios and other alternatives}
\hspace*{\parindent}
In this section we make some comments on the inverted scenarios and other
alternatives of the above idea for Affleck-Dine baryogenesis.
As in the models for hybrid inflaton\cite{hybrid_original},
we can construct inverted scenarios\cite{inverted} for our model.
Perhaps the easiest way is;

1) Consider a source brane where a source field for the supersymmetry
breaking is localized.
In the true vacuum, this brane is located at a distance from the
``SM-brane'' where standard model fields are located.

2) The trigger field is the moduli for the distance between the source
   brane and the SM-brane. The potential for the moduli is destabilized
   in the true vacuum by the $O(m_{3/2})$ correction, which makes these
   branes separated in the true vacuum.

3) During inflation and the follwing stage of the oscillation, the
   $O(H)>m_{3/2}$ correction stabilizes the moduli. Thus the branes stay
   on top of each other during this period.

4) When branes are located on top of each other, the corrections from the
   source brane destabilize some of the flat directions of the
   MSSM. During this period,  the condensate of the Affleck-Dine field
   is expected.

5) As the Hubble parameter becomes small, the trigger field (the moduli
   for the brane distance) is destabilized. Then the branes move toward
   their minima. At this stage, the Affleck-Dine field is stabilized to
   start oscillation.

The model is quite similar to the original model that we have stated in
the previous section, except for the
mechanism for the supersymmetry breaking.

An alternative scenario for the inverted model is already discussed
in ref.\cite{ADafterThermal}, in which Affleck-Dine baryogenesis after
thermal brane inflation is discussed.

In ref.\cite{matsuda_nontach}, a modified version of thermal brane
inflation is discussed.
As the thermal hybrid inflation starts at a distance, one 
can easily construct ``not inverted''
scenario for Affleck-Dine baryogenesis after thermal brane
inflation.

\section{Conclusions and Discussions}
\hspace*{\parindent}
At first, there had not been no concrete mechanism to arrange the suitable
initial condition for the Affleck-Dine mechanism.
Later observations\cite{OH-corrections} showed that the gravitational 
corrections of $O(H)$ can
destabilize the flat direction to yield the expected initial condition
for the mechanism.
However, some later discussions suggested\cite{Q-ball_kusenko,
Problem_of_Q-ball}  
that this simple mechanism
cannot work in its simplest form because of the non-trivial formation of
the Q-balls or the thermal effects.
Thus the succeeding scenarios are strongly model dependent.
For models with large extra dimensions, the problem
is quite serious for the original single-field models.

In this paper we extend the original single-field models to include an
additional trigger field.
The result is quite favorable for models with large extra dimensions. 
We have also constructed inverted scenarios and other alternatives.

\section{Acknowledgment}
We wish to thank K.Shima for encouragement, and our colleagues in
Tokyo University for their kind hospitality.

\end{document}